\PassOptionsToPackage{square,numbers}{natbib}
\documentclass[aps,nofootinbib,10pt]{revtex4-2}
\usepackage[T1]{fontenc}
\usepackage[utf8]{inputenc}
\usepackage[english]{babel}
\usepackage{graphicx} 
\usepackage{amsfonts}
\usepackage{amsmath}
\usepackage{amsthm}
\usepackage{subfig}
\usepackage[title]{appendix}
\usepackage{amssymb}
\usepackage{tensor}
\usepackage{hyperref}
\usepackage{stmaryrd}
\hypersetup{
    colorlinks,
    citecolor=red,
    linkcolor=blue,
    urlcolor=red
}
\usepackage{empheq}
\usepackage{enumitem}
\usepackage[compress]{cleveref}
\crefrangelabelformat{equation}{(#3#1#4--#5\crefstripprefix{#1}#2#6)}

\newcommand{\I}{\indices}
    \newcommand{\p}{^{(\phi)}}

    \newcommand{\apr}{\dot{a}}
    \newcommand{\as}{\ddot{a}}
 
    \newcommand{\alp}{\dot{\alpha}}
    \newcommand{\als}{\ddot{\alpha}}
    \newcommand{\bp}{\dot{\beta}}
    \newcommand{\bs}{\ddot{\beta}}
    \newcommand{\gp}{\dot{\gamma}}
    \newcommand{\gs}{\ddot{\gamma}}
    \newcommand{\Mp}{M_{P}}
    \newcommand{\Mpd}{M_{P}^2}
    
\begin{document}

\title{Cosmological scenarios from topological invariants: Inflation and Quintom theory with Barbero-Immirzi scalar field.}

\author{Pollari Giacomo} 

\thanks{Electronic address: 504077@mail.muni.cz}

	\affiliation{Department of Theoretical Physics and Astrophysics,
		Faculty of Science of the Masaryk University,
		Kotlářská 2, 611 37 Brno, Czech Republic}

\maketitle

\begin{center}
\textbf{Abstract} 
\end{center}
    In the present work we provide a cosmological description of the Palatini action with the addition of three well-known topological invariants, each of which is attached to a scalar field. As a consequence, torsion plays a central role, yielding an effective action with three dynamical scalar fields coupled to gravity subjected to the influence of a potential. A special heed will be addressed to inflation scenarios and quintom theory, although other noteworthy cases will be analyzed. Two out of three scalars can be thought as the Barbero-Immirzi scalar field as coupled to the Holst and Nieh-Yan terms with a potential whereby they shape the evolution of the universe. The other field turns out to be a phantom that we couple to the inflaton that at late times holds the stage inducing a fast expansion.

	\section{Introduction} \noindent
	The theory of General Relativity (GR) was proven to be the most accurate description of gravity \citep{Wald} \citep{Thorne}. There are two equivalent ways of writing the action of pure gravity: \textit{first} and \textit{second order} formalisms. The latter corresponds to the Einstein-Hilbert (EH) action whose connection depends on the metric tensor and is Levi-Civita, i.e. torsion free. The first is the Palatini (P) action where the connection is considered to be independent from the metric. The variation of the P action with respect to the connection leads to the torsion-free condition and both the EH and the P actions give the same Einstein field equations. However, the equations of motion remain unchanged also by adding topological invariants of the curvature and/or the torsion. The first  discovered was the Gauss-Bonnet-Chern \citep{Chern1},\citep{Chern2} giving the Euler characteristic of the manifold. Another topological piece not involved in the dynamics is the Pontryagin density which is related to the Chern-Simons current for GR \citep{Pontryagin1} \citep{Pontryagin2}. Subsequently a further one that joined the family was found in 1982 by Nieh and Yan \citep{Nieh} which stemmed from the generalized Bianchi identities. \\ \noindent
	For decades the quantization of gravity and its coupling with matter fields of the SM has been the scope of Quantum Gravity. Loop Quantum Gravity (LQG) is one of the most well-developed attempts constructed from GR that expresses gravity as a Yang-Mills SU(2) gauge theory \citep{Thiemann} \citep{Rovelli} \citep{Gambini} \citep{Gambini2} \citep{CQG}. In 1996 S\"{o}ren Holst proved that there exists another topological invariant in the Palatini formulation \citep{Holst} from which the Ashtekar-Barbero connection naturally arises from the action. This allows the GR action to be SU(2) invariant as a consequence of the presence of the Gauss constraint in the Hamiltonian picture. From such a Holst invariant the Immirzi parameter $\gamma:= 1/\beta$ emerges as a parameter, entering the value of the area operator spectrum \citep{Rovelli} \citep{Rovelli2} $\textbf{A}_{\vec{j}} = 8\pi \gamma \hslash G c^{-3} \sum_{i} \sqrt{j_i\left(j_i+1\right)}$, therefore its meaning is crucial both in a topological and in a more pragmatic way. The first value obtained $\gamma= ln2 /\pi \sqrt{3}$ was found by Ashtekar \textit{et al.} \citep{AshtekarImm} by comparing the classical Bekerstein-Hawking entropy of black holes to the quantum area of LQG. A further study for different types of black hole horizons suggested slight different values \citep{AshtekarImm2}. However, the debate around it still continues and it seems that the Immirzi parameter might not be fixed to remove any ambiguity. Another possibility taken into account in \citep{Taveras} \citep{CM1} \citep{Mercuri1} was to promote the Immirzi parameter to a real scalar field both in the Holst and Nieh-Yan action and obtain the effective action which turns out to be a scalar-tensor of the Immirzi field minimally coupled to gravity. \\ \noindent
	The presence of these terms in pure gravity does not affect the equations of motion, yet the addition of matter fields such as fermions makes the first and second order formulations differ \citep{Rovelli} \citep{Rovelli2}. Therefore all the topological invariants have to be looked at in a different light and can play a crucial role in the dynamics of the gravity-matter coupling. \\ \noindent
	In addition to the ones mentioned above we also consider a further piece that for simplicity we call Nieh-Yan-like topological invariant that we shall show later. Unlike the original Nieh-Yan, it cannot be written in terms of curvature and torsion in the first order formulation, but it emerges as a part of the spin curvature when the spin connection is split as a sum of the Lorentz one and the contorsion. A natural application of scalar fields in Modified Gravity is undoubtedly in cosmological models. In particular, the theory of inflation \citep{Guth1} \citep{Guth2} \citep{Liddle-Lyth} \citep{CI1} \citep{CI2} \citep{Copeland} is one of the most astounding and prominent solution to a lot of problems of the old theory of the Big Bang, such as the flatness, horizon, monopole problems and so forth. It attempts to explain what could have occurred in the very early universe and how matter was created at the end of it. Furthermore, not only a scalar field is invoked to describe the beginning of everything but also the later stages, e.g. the accelerated universe we are experiencing. To this purpose an exotic scalar field known as the phantom (or ghost) is an hypothetical particle that is thought to be responsible for the expansion after a matter dominated universe, specially as an alternative to Dark Energy \citep{Odintsov} \citep{Sami} \citep{PhantomCosmologies} \citep{Doom} \citep{Ods} \citep{Infl-Phan}. A theory that combines both the fields is called Quintom \citep{Quin1} \citep{Quin2} \citep{Quin3}. As a cosmological implementation of our results we shall put a particular attention on inflation and quintom theory even though other noteworthy cases will be studied. \\ \noindent	
	In the following we will use the convention under which the Levi-Civita connection $\bar{\Gamma} \I{_\mu _\nu ^\rho} = \Gamma \I{_{(\mu}_{\nu)}^\rho} $ is symmetric in the first two indices and the torsion $T \I{_\mu _\nu ^\rho} = 2\Gamma \I{_{[\mu}_{\nu]}^\rho}$ is skew-symmetric in the first two, while the contorsion tensor is antisymmetric in the last two indices $K \I{_\mu _\nu_\rho} = - K \I{_\mu _\rho_\nu}$. Greek letters $\mu,\nu,\ldots =0,1,2,3$ are used for spacetime indices and Latin $I,J,\ldots = 0,1,2,3$ for Lorentz (flat) ones. The signature is $(-,+,+,+)$. Since the metric tensor can be written in an orthonormal basis via the tetrads $g_{\mu \nu} = e_\mu^I e_\nu^J \eta_{IJ}$, in order to avoid confusion, we denote as $\nabla_\mu$ a $g$-compatible affine connection and with $D_\mu$ the connection compatible with the tetrads, i.e. $D_\mu e_\nu^I = \nabla_\mu e_\nu^I + \omega \I{_\mu^I_J} e_\nu^J =0$ where $\omega \I{_\mu^I_J}$ is the spin connection antisymmetric in the last two indices.
\\ \noindent
	This paper is structured as follows. \\ \noindent
   In Section II we review the Nieh-Yan and Nieh-Yan-like four-form written in terms of the irreducible torsion components, how they can be decomposed and how they change under a conformal transformation. \\ \noindent
   In Section III we first consider the Palatini action with topological invariants, select the Holst, Nieh-Yan and Nieh-Yan-like, promote their coefficients to scalar fields and solve for the irreducible component of the torsion. Secondly, we write the effective action adding a potential and obtain the equations of motion. \\ \noindent
   In Section IV we obtain the EOM in flat FLRW and divide the work into two specific cases: the first one involve only the Barbero-Immirzi scalar fields with further subcases of three given potentials, the second one is a classical example of quintom theory where one field plays the role of the inflaton and the other the phantom with an interacting potential between them. \\ \noindent
   In Section V we draw the conclusions of our work highlighting the interesting and also problematic outcomes.
   
 \section{Review of some topological invariants} \label{Sec.2}
	The Nieh-Yan 4-form given by
	\begin{equation}\label{2.1}
	    d\left( e^I \wedge T_I \right) = T^I \wedge T_I - e^I \wedge e^J \wedge R \I{_I_J}
	\end{equation}
	by Stokes theorem reduces to a surface integral that vanishes on the boundary. Writing \eqref{2.1} in coordinate basis:
    $$ d\left( e^I \wedge T_I \right) = \dfrac{1}{2} d\left( e^I_\nu T \I{_\rho_\sigma_I} dx^\nu \wedge dx^\rho \wedge dx^\sigma \right) = \dfrac{1}{2} \left( e^I_\nu T \I{_\rho_\sigma_I} \right)_{,\mu}  dx^\mu \wedge dx^\nu \wedge dx^\rho \wedge dx^\sigma$$
    leads to
    \begin{equation}\label{2.2}
       d\left( e^I \wedge T_I \right) = \dfrac{1}{2}\left( e^I_\nu T \I{_\rho_\sigma_I} \right)_{,\mu} \epsilon\I{^\mu^\nu^\rho^\sigma} d^4x
    \end{equation}
    where, $ (T_I)_{\mu \nu}:= \dfrac{1}{2} T_{\mu \nu I} $ and  $\epsilon^{\mu \nu \rho \sigma}$ is the Levi-Civita symbol, related to the Levi-Civita tensor $\varepsilon_{\mu \nu \rho \sigma}$ by $\varepsilon_{\mu \nu \rho \sigma} = \sqrt{-g} \epsilon_{\mu \nu \rho \sigma}$ and $\varepsilon^{\mu \nu \rho \sigma} = \epsilon^{\mu \nu \rho \sigma} /\sqrt{-g}$. Thus, contracting the Lorentz indices and integrating over all the spacetime $\mathcal{M}$ we obtain
    \begin{equation}\label{2.3}
        \int_\mathcal{M} d\left( e^I \wedge T_I \right) = \dfrac{1}{2}\int_\mathcal{M} d^4x \left(T \I{_\rho_\sigma_\nu} \right)_{,\mu} \epsilon\I{^\mu^\nu^\rho^\sigma}.
    \end{equation}
    Since the partial derivative of the Levi-Civita symbol vanishes, then
    \begin{equation}\label{2.4}
        \int_\mathcal{M}  d\left( e^I \wedge T_I \right) = \dfrac{1}{2}\int_\mathcal{M} d^4x \partial_\mu \left(T \I{_\rho_\sigma_\nu} \epsilon\I{^\mu^\nu^\rho^\sigma} \right) = \dfrac{1}{2}\int_\mathcal{M} d^4x \partial_\mu \left( \sqrt{-g}T \I{_\rho_\sigma_\nu} \varepsilon\I{^\mu^\nu^\rho^\sigma} \right)
        \end{equation}
    which vanishes iff 
    $$\left(T \I{_\rho_\sigma_\nu} \varepsilon\I{^\mu^\nu^\rho^\sigma}\right)|_{\partial \mathcal{M}} =0 .$$
    It is useful to decompose the torsion tensor in its irreducible components according to the Lorentz group \citep{Shapiro} \citep{CM1} \citep{Capozziello}:
    \begin{equation}\label{2.5}
        T_{\mu \nu \rho} = \dfrac{1}{3} \left( T_\nu g_{\mu \rho} - T_\mu g_{\nu \rho} \right) - \dfrac{1}{6} \varepsilon \I{_\mu_\nu_\rho_\sigma} S^\sigma + q \I{_\mu_\nu_\rho}
    \end{equation}
    where $T^\mu = T \I{^\nu^\mu_\nu}$ is the \textit{trace vector} and carries 4 DOF, $S^\sigma$ is the \textit{(pseudotrace) axial vector} with 4 DOF and $q \I{_\mu_\nu_\rho}$ is the non totally skew-symmetric traceless part of the torsion that has the remaining 16 DOF and such that $\epsilon \I{^\mu^\nu^\rho^\sigma} q_{\mu \nu \rho} =0$. \\ \noindent
    Inserting \eqref{2.5} into \eqref{2.4} and contracting, it gives
    \begin{equation}\label{2.6}
        \int_\mathcal{M} d\left( e^I \wedge T_I \right) = \dfrac{1}{2} \int_\mathcal{M} d^4x \sqrt{-g} \bar{\nabla}_\mu S^\mu
        \end{equation}
    which agrees with \citep{CM1}. Therefore only the axial vector has to vanish on the boundary and not all the other irreducible components. Following the same concepts, if 
    $$ \int_\mathcal{M} d\left( e^I \wedge T_I \right) $$
    by Stokes theorem vanishes because some of the components of $T_I$ do on the boundary, so has to do the four form
    \begin{equation}\label{2.7}
    d\left(e^I \wedge \star T_I\right)
    \end{equation}
    that we dub Nieh-Yan-like, where $\star$ is the Hodge operator. Indeed \eqref{2.7} is the only other non trivial exact four form that can be constructed from the torsion. Such a piece is studied in teleparallel gravity \citep{TP1} \citep{TP2}. Expressing the torsion 2-form in holonomic basis and then apply the Hodge operator we have
	$$\star T_I =\dfrac{1}{2} T\I{_\alpha_\beta_I} \star \left(dx^\alpha \wedge dx^\beta \right) = \dfrac{1}{4} \sqrt{-g} T\I{_\alpha_\beta_I} g^{\alpha \gamma} g^{\beta \delta} \epsilon\I{_\gamma_\delta_\rho_\sigma} dx^\rho \wedge dx^\sigma $$
    which put into \eqref{2.7} gives
    \begin{equation}\label{2.8}
    d\left(e^I \wedge \star T_I\right) = \dfrac{1}{4} \left(\sqrt{-g} e_\nu^I T\I{_\alpha_\beta_I} g^{\alpha \gamma} g^{\beta \delta} \epsilon\I{_\gamma_\delta_\rho_\sigma}\right)_{,\mu} dx^\mu \wedge dx^\nu \wedge dx^\rho \wedge dx^\sigma =\dfrac{1}{4} \left( \sqrt{-g} T\I{_\alpha_\beta_\nu} g^{\alpha \gamma} g^{\beta \delta} \epsilon\I{_\gamma_\delta_\rho_\sigma}\right)_{,\mu} \epsilon^{\mu \nu \rho \sigma} d^4x.
    \end{equation}
   Thus, eq.\eqref{2.8} becomes
    $$d\left(e^I \wedge \star T_I\right) = \dfrac{1}{4}  \left( \sqrt{-g} T\I{^\gamma^\delta_\nu} \epsilon\I{_\gamma_\delta_\rho_\sigma}\right)_{,\mu} \epsilon^{\mu \nu \rho \sigma} d^4x = \dfrac{1}{4} \sqrt{-g} \: \left(\bar{\nabla}_\mu T\I{^\gamma^\delta_\nu}  \epsilon\I{_\gamma_\delta_\rho_\sigma}\epsilon^{\mu \nu \rho \sigma} \right) d^4x.$$
    Contracting all the quantities and using \eqref{2.5} we obtain
    \begin{equation}\label{2.9}
    	d\left(e^I \wedge \star T_I\right) = \sqrt{-g} \: \bar{\nabla}_\mu T\I{^\mu} d^4x.
    \end{equation}
    Therefore, its integral
    \begin{equation}\label{3.0}
    	\int_\mathcal{M} d\left(e^I \wedge \star T_I\right) = \int_\mathcal{M} d^4x \;\sqrt{-g} \, \bar{\nabla}_\mu T\I{^\mu}= \int_\mathcal{M} d^4x \; \partial_\mu \left(\sqrt{-g} \, T\I{^\mu} \right)
    \end{equation}
    by the divergence theorem reduces to a surface integral that vanishes iff $T\I{^\mu}|_\mathcal{M}=0$. In the Nieh-Yan case the totally antisymmetric part of the torsion, i.e. the axial vector, has to vanish whereas in the Nieh-Yan-lika its trace part.
  
  \subsection{Decomposition}
  The RHS of \eqref{2.1} is equivalent to its LHS by using the Cartan structure equations. Instead, the boundary term \eqref{2.7} cannot be a combination of the torsion and the curvature maintaining the connection independent. First, we write the action
  \begin{equation}\label{4.0}
      S = \int_\mathcal{M} d^4x \; e \: e^\mu_I e^\nu_J R \I{_\mu_\nu^I^J}
  \end{equation}
  decompose the spin connection $\omega \I{_\mu^I^J} = \bar{\omega} \I{_\mu^I^J} - K \I{_\mu^I^J}$:
  \begin{equation}\label{4.1}
  	 S = \int_\mathcal{M} d^4x \; e \,\left[ e^\mu_I e^\nu_J \bar{R} \I{_\mu_\nu^I^J} - 2 e^\mu_I e^\nu_J \bar{D}_\mu K \I{_\nu^I^J} + e^\mu_I e^\nu_J \left( K \I{_\mu^I^N} K \I{_\nu_N^J} - K \I{_\nu^I^N} K \I{_\mu_N^J} \right)  \right]
  \end{equation}
  where $\bar{D}_\mu$ is the Levi-Civita tetrad-compatible connection. Contracting the second term of \eqref{4.1} and using \eqref{2.5} we end up with
  \begin{equation}\label{4.2}
  	 S = \int_\mathcal{M} d^4x \; e \,\left[ e^\mu_I e^\nu_J \bar{R} \I{_\mu_\nu^I^J} - 2 \bar{\nabla}_\mu T^\mu + \left( K \I{_\mu^\mu^\rho} K \I{_\nu_\rho^\nu} - K \I{_\nu^\mu^\rho} K \I{_\mu_\rho^\nu} \right)  \right].
  \end{equation}
  From \eqref{4.2} we see that the integral of \eqref{2.7} corresponds to the Levi-Civita covariant derivative of the trace vector coming from the decomposition of the connection. This is not surprising whatsoever. The integral of the Nieh-Yan form \eqref{2.1}, following the same procedure we have just shown, leads to \eqref{2.6}. In particular, the contorsion terms of the splitting of the connection in the curvature cancel the torsion-torsion part and the curvature constructed from the Levi-Civita connection vanishes by virtue of the Bianchi identity. Therefore, in that case the only surviving term is exactly \eqref{2.6}. However, in \eqref{4.2} the presence of the curvature does not allow to recast $d(e^I \wedge \star T_I)$ in a way in which we have a curvature with a connection independent from the metric. From \eqref{4.2} we can single out  the topological term and write
  \begin{equation}\label{4.3}
  	  2d(e^I \wedge \star T_I) = \star \left(e^I \wedge e^J\right) \wedge \left( \bar{R}_{JI} - {R}_{JI} + K \I{_I^K} \wedge K \I{_K_J} \right)= \star \left(e^I \wedge e^J\right) \wedge \bar{D}K_{IJ}
  \end{equation}
  where
  $$ \bar{D}K_{IJ} = dK_{IJ} + 2\bar{\omega}\I{_{[I|}^K} \wedge K \I{_K_{|J]}}$$
  is the Levi-Civita covariant derivative of the contorsion with respect to a non-coordinate basis. Equivalently, the action \eqref{4.0} can also be rewritten in terms of forms as 
  \begin{equation}\label{4.4}
  	  S = \int_{\mathcal{M}} e^I \wedge e^J \wedge \star R_{JI} = \int_{\mathcal{M}} \left[\star \left( e^I \wedge e^J\right) \wedge \left(\bar{R}_{JI} + K \I{_I^K} \wedge K \I{_K_J}\right) - 2 d(e^I \wedge \star T_I) \right]
  \end{equation}
  Clearly \eqref{2.7} is Diff($\mathcal{M}$) and SO(1,3) invariant. We would like to know how it transforms under a conformal transformation and how \eqref{4.0} changes. In the Cartan structure equations the spin connection is a gauge field, treated as an independent variable that does not change under any transformation of the metric. In accordance with \citep{Shapiro}, under the conformal transformation 
  \begin{equation}\label{4.5}
      \tilde{g}_{\mu \nu}(x) = \phi^2(x) g_{\mu \nu}(x) \hspace{2cm} \tilde{e}^I(x) = \phi(x) e^I(x) ,
  \end{equation}
  with $\phi \in \Omega^0(\mathcal{M})$, the first Cartan structure equation becomes
    $$ \tilde{T}^I = d\tilde{e}^I + \omega \I{^I_J} \wedge \tilde{e}^J$$
    from which we obtain
    \begin{equation}\label{4.6}
        \tilde{T}^I = \phi T^I + d\phi \wedge e^I.
    \end{equation}
  It is easy to see that by expressing \eqref{4.6} in components we have
  $$ \tilde{T} \I{_\mu_\nu^\rho} = \tilde{T} \I{_\mu_\nu^I} \tilde{e}^\rho_I = T \I{_\mu_\nu^\rho} + \dfrac{1}{\phi}\left( \phi_{,\mu} \delta_\nu^\rho - \phi_{,\nu} \delta_\mu^\rho \right)$$
  and by using the decomposition \eqref{2.5}
  the trace vector changes as
  \begin{equation}\label{4.7}
      \tilde{T}_\mu = T_\mu - \dfrac{3}{\phi}\phi_{,\mu}.
  \end{equation}
  By virtue of this we can write the conformally transformed term \eqref{2.7} as
  $$ d(\tilde{e}^I \wedge \star \tilde{T}_I) = d(\phi e^I \wedge \star ( \phi T_I + d\phi \wedge e_I))$$
  $$ = \dfrac{1}{4} d \left(\sqrt{-g} \phi e_\nu^I g^{\alpha \gamma} g^{\beta \delta} \varepsilon_{\alpha \beta \rho \sigma} \left(\phi T_{\gamma \delta I} + (\phi_{,\gamma} e_{\delta I} - \phi_{,\delta} e_{\gamma I}) \right) \right) dx^\nu \wedge dx^\rho \wedge dx^\sigma$$
    $$ -\dfrac{1}{2} \left(\sqrt{-g} \phi g^{\alpha \gamma} g^{\beta \delta} \left( \phi T_{\gamma \delta \nu} + \phi_{,\gamma} g_{\delta \nu} - \phi_{,\delta} g_{\gamma \nu}\right)\right)_{,\mu} (\delta^\mu_\alpha \delta^\nu_\beta - \delta^\mu_\beta \delta^\nu_\alpha)  d^4x,$$
    using \eqref{2.5} and contracting it yields
    \begin{equation}\label{4.8}
        d(\tilde{e}^I \wedge \star \tilde{T}_I) = \sqrt{-g} \bar{\nabla}_\mu \left( \phi^2 T^\mu - 3\phi \phi \I{_,^\mu} \right) d^4x = \sqrt{-g} \bar{\nabla}_\mu (\phi^2 \tilde{T}^\mu) d^4x
    \end{equation}
    where in the last line we used \eqref{4.7}. It is not surprising that \eqref{4.8} is again a boundary term. Now we want to derive \eqref{4.0} under a conformal transformation \eqref{4.5} in presence of torsion. We denote the torsion-free $\tilde{e}$-compatible connection as 
    \begin{equation}\label{4.9}
        \tilde{D}_\mu \tilde{e}_\nu^I = \tilde{e}^I_{\nu,\mu} - \tilde{\Gamma} \I{_\mu_\nu^\rho} \tilde{e}^I_\rho + \tilde{\omega} \I{_\mu^I_J} \tilde{e}^J_\nu =0
    \end{equation}
    where $\tilde{\Gamma} \I{_\mu_\nu^\rho} = \bar{\Gamma} \I{_\mu_\nu^\rho} + \tensor{\p\Gamma}{_\mu_\nu^\rho}$ and $\tilde{\omega} \I{_\mu^I^J} = \bar{\omega} \I{_\mu^I^J} + \tensor{\p\omega}{_\mu^I^J}$, $\tensor{\p\Gamma}{_\mu_\nu^\rho}$ and $\tensor{\p\omega}{_\mu^I^J}$ being the part of the connection containing the derivative of $\phi$. The conformally transformed action \eqref{4.0} is
    $$  \tilde{S} = \int_\mathcal{M} d^4x\; \tilde{e} \; \tilde{e}^\mu_I \tilde{e}^\nu_J \tilde{R} \I{_\mu_\nu^I^J} $$
    $$= \int_\mathcal{M} d^4x\; e \; \phi^2 e^\mu_I e^\nu_J \left[ 2\tilde{\omega}_{[\nu,\mu]} - 2\tilde{K}_{[\nu,\mu]} + 2 \left(\tilde{\omega} \I{_{[\mu}^I^K} - \tilde{K} \I{_{[\mu}^I^K}) \right) \left(\tilde{\omega} \I{_{\nu]}_K^J} - \tilde{K} \I{_{\nu]}_K^J}) \right) \right]$$
    where $\tilde{K} \I{_\nu^I^J}$ is the transformed contorsion tensor. Rearranging the terms it yields
    \begin{equation}\label{4.10}
        \tilde{S} = \int_\mathcal{M} d^4x\; e \; \phi^2 e^\mu_I e^\nu_J \left( \tensor{\p\tilde{R}}{_\mu_\nu^I^J} - 2 \tilde{D}_{\mu} \tilde{K} \I{_{\nu}^I^J} + 2 \tilde{K} \I{_{[\mu}^I^K} \tilde{K} \I{_{\nu]}_K^J} \right)
    \end{equation} 
    with
    $$ \tensor{\p\tilde{R}}{_\mu_\nu^I^J} = 2\tilde{\omega}_{[\nu,\mu]} +2 \tilde{\omega} \I{_{[\mu}^I^K} \tilde{\omega} \I{_{\nu]}_K^J}.$$
    Using the compatibility of the connection \eqref{4.9} and that the trace of the contorsion equals \eqref{4.7}, the second term in \eqref{4.10} gives
    $$ -2 \phi^2 e^\mu_I e^\nu_J \tilde{D}_{\mu} \tilde{K} \I{_{\nu}^I^J} = -2 \phi^2 \tilde{\nabla}_{\mu} \tilde{T}^\mu +4 \phi \phi_{,\mu} \tilde{T}^\mu, $$
    and expressing the connection $\tilde{\nabla}_\mu$ on the trace vector in terms of $\bar{\nabla}_\mu$ it yields
    $$ \tilde{\nabla}_\mu \tilde{T}^\mu = \bar{\nabla}_\mu \tilde{T}^\mu + \dfrac{4}{\phi} \phi_{,\mu}\tilde{T}^\nu.$$
    Therefore the action \eqref{4.10} becomes 
    \begin{equation}\label{4.11}
        \tilde{S} = \int_\mathcal{M} d^4x\; e \; \left(\phi^2 e^\mu_I e^\nu_J \tensor{\p\tilde{R}}{_\mu_\nu^I^J}- 2 \bar{\nabla}_\mu \big(\phi^2 \tilde{T}^\mu\big) + 2 \phi^2 e^\mu_I e^\nu_J\tilde{K} \I{_{[\mu}^I^K} \tilde{K} \I{_{\nu]}_K^J} \right).
    \end{equation}
    We see that the second term in brackets of \eqref{4.11} coincides (up to a factor 2) with \eqref{4.8}. In our case we used the decomposition of the torsion tensor in its irreducible components but the equality mentioned holds also for the general undecomposed one. Hence, this allows to write the conformally transformed action in the language of forms as
    \begin{equation}\label{eq:4.11}
        \tilde{S} = \int_{\mathcal{M}} \tilde{e}^I \wedge \tilde{e}^J \wedge \star \tilde{R}_{JI} = \int_{\mathcal{M}} \left[\star \left( \tilde{e}^I \wedge \tilde{e}^J\right) \wedge \left(\tensor{\p\tilde{R}}{_J_I} + \tilde{K} \I{_I^K} \wedge \tilde{K} \I{_K_J}\right) - 2 d(\tilde{e}^I \wedge \star \tilde{T}_I) \right]
    \end{equation}
    thus, it is the same as just transforming every quantity in the action \eqref{4.4} under \eqref{4.5} while preserving its form. It can be shown that the same result holds for \eqref{2.1}. 

    \section{Topological invariants coupled to scalar fields} \label{Sec.3}
    We can write down the Palatini action together with all the topological invariants mentioned in the Introduction. In natural units, it reads 
    \begin{equation} \label{5.0}
    \small
    S = \dfrac{\Mpd}{2}\int \Bigl( \underbrace{e^I \wedge e^J \wedge \star R_{JI}}_{PALATINI} + \underbrace{\alpha_1 e^I \wedge e^J \wedge R_{IJ}}_{HOLST} + \underbrace{\alpha_2 R^{IJ} \wedge R_{JI}}_{PONTRYAGIN} + \underbrace{\alpha_3 R^{IJ} \wedge \star R_{JI}}_{GAUSS-BONNET} + \underbrace{\alpha_4 d(e^I \wedge T_I)}_{NIEH-YAN} + \underbrace{\alpha_5 d(e^I \wedge \star T_I)}_{NIEH-YAN-LIKE} \Bigr)
    \end{equation}
    where $\{\alpha_i\}_{i=1}^5$ are arbitrary dimensionless constants and $\Mp$ is the reduced Planck mass. If we promote $\{\alpha_i\}_{i=1}^5$ to scalar fields, these five terms are no longer irrelevant as torsion emerges from the variation of the action with respect to the connection. In the following we focus on the Holst, Nieh-Yan and Nieh-Yan-like in \eqref{5.0} and drop the Pontryagin and Bonnet terms. Now let us promote the coefficients $\alpha_1,\alpha_4,\alpha_5$ to scalar fields and study the action  
    $$ S= \dfrac{\Mp}{2}\int \Bigl( \Mp e^I \wedge e^J \wedge \star R_{JI} + \alpha(x) e^I \wedge e^J \wedge R_{IJ}+ \beta(x) d(e^I \wedge T_I) + \gamma(x) d(e^I \wedge \star T_I) \Bigr)$$
    with $\alpha(x),\beta(x),\gamma(x)$ arbitrary scalar fields (either real or complex). We rescaled them by $\Mp$, i.e. $\Mp \alpha := \alpha$, so that they have the correct physical dimensions in natural units \footnote{It is perfectly legit to rescale them by a constant since scalar fields are not physical observables.}. Instead of varying the action with respect to $\omega\I{_\mu^I_J}$ which has 24 DOF, we decompose the connection as $\omega \I{_\mu^I_J} = \bar{\omega} \I{_\mu^I_J} + K \I{_\mu^I_J}$ and vary with respect to the 24 irreducible components of the torsion. Using \eqref{2.6}, \eqref{3.0}  and after some calculation on the Palatini and Holst parts, we obtain the following action 
    \begin{equation}\label{5.1}
S= \dfrac{\Mp}{2}\int_{\mathcal{M}} d^4x \; e\; \left[ \Mp e^\mu_I e^\nu_J \bar{R} \I{_\mu_\nu^I^J} 
-2\partial_\mu \gamma T^\mu -\dfrac{1}{2} \left(\partial_\mu \alpha + \partial_\mu \beta\right) S^\mu+\dfrac{\alpha}{3} T_\mu S^\mu -\dfrac{2}{3}T_\mu T^\mu +\dfrac{1}{24} S_\mu S^\mu +\dfrac{1}{2}q_{\mu \nu \rho} q^{\mu \nu \rho}\right].
\end{equation} 
Solving for the irreducible components $\{T^\mu,S^\mu,q^{\mu \nu \rho}\}$, it yields
\begin{subequations} 
    \begin{empheq}[left=\empheqlbrace]{align}
 &T_\mu = \dfrac{3}{2} \dfrac{1}{\alpha^2 + \Mpd} \left[ \alpha(\partial_\mu \alpha + \partial_\mu \beta) - \Mp \partial_\mu \gamma\right] \label{5.2a} \\
 &S_\mu = \dfrac{6}{\alpha^2+\Mpd} \left[\Mp(\partial_\mu \alpha +\partial_\mu \beta) + \alpha \partial_\mu \gamma \right] \label{5.2b}\\
&q_{\mu \nu \rho} =0 \label{5.2c}
\end{empheq}
\end{subequations}
Inserting \crefrange{5.2a}{5.2c} into \eqref{5.1}, the effective action reads
\begin{equation}\label{5.3}
\small
    S= \dfrac{\Mpd}{2}\int_{\mathcal{M}} d^4x \; e\;  e^\mu_I e^\nu_J \bar{R} \I{_\mu_\nu^I^J} - \dfrac{3}{4} \int_{\mathcal{M}} d^4x \; e\; \dfrac{\Mpd}{\alpha^2+\Mpd}\left[ (\partial_\mu\alpha + \partial_\mu\beta)(\partial^\mu\alpha  + \partial^\mu \beta) + \dfrac{2\alpha}{\Mp}(\partial_\mu \alpha + \partial_\mu \beta) \partial^\mu \gamma - \partial_\mu \gamma \partial^\mu \gamma\right].
\end{equation}
It is worth noting that for $\alpha,\gamma=0$ and $\beta,\gamma=0$ it reduces to the Holst and Nieh-Yan case of \citep{CM1}, respectively. However we got an additional field from the trace part of the torsion that was not present in previous works. Since the kinetic term is non canonical and there exists no redefinition of the scalar fields such that we end up with a standard kinetic part, in order to have a real Lagrangian, $\alpha,\beta,\gamma$ could be complex fields. In particular $\alpha^2+\Mpd$ in the denominator is forced to be real, so $\alpha$ is either real or imaginary and as a consequence $\beta$ can in general be complex. If $\alpha$ is real and $\beta$ is complex such that $\partial_\mu\alpha + \partial_\mu\beta$  is real, $\gamma$ is also real. If $\alpha$ is imaginary and $\beta$ is complex such that $\partial_\mu\alpha + \partial_\mu\beta$ is real, then $\gamma$ is imaginary. This non-canonical kinetic energy can be thought as a manifestation of interaction among the fields whose role of mediator is exerted by $\alpha$. The only requirement we impose is that $\alpha\neq \pm i\Mp$. In addition to it we add a potential in \eqref{5.3}:
\begin{equation}\label{5.4} \small
    S= \int_{\mathcal{M}} d^4x \; \sqrt{-g}\; \left[\dfrac{\Mpd}{2} R - \dfrac{3}{4} \dfrac{\Mpd}{\alpha^2+\Mpd}\left[ (\partial_\mu\alpha + \partial_\mu\beta)(\partial^\mu\alpha  + \partial^\mu \beta) + \dfrac{2\alpha}{\Mp}(\partial_\mu \alpha + \partial_\mu \beta) \partial^\mu \gamma - \partial_\mu \gamma \partial^\mu \gamma\right] -V(\alpha,\beta,\gamma)\right]
\end{equation}
hence falling into a scalar-tensor theory with three scalar fields minimally coupled to gravity with a nonstandard kinetic term. Solving the Euler-Lagrangian equations for $\alpha,\beta,\gamma$ and the metric, we arrive at the system
\begin{subequations} \label{5.5} \small
    \begin{align} 
 &\alpha: \hspace{0.08cm}  - \alpha (\partial_\mu \alpha \partial^\mu \alpha - \partial_\mu \beta \partial^\mu \beta + \partial_\mu \gamma \partial^\mu \gamma) + \dfrac{\alpha^2 - \Mpd}{\Mp}\partial_\mu \beta \partial^\mu \gamma + (\alpha^2+\Mpd) \left( \bar\Box\alpha + \bar\Box\beta + \dfrac{\alpha}{\Mp} \bar\Box\gamma\right) - \dfrac{2(\alpha^2+\Mpd)^2}{3\Mpd}V_{,\alpha}=0 \label{5.5a} \\
 &\beta: \hspace{0.3cm}  2\alpha (\partial_\mu \alpha \partial^\mu \alpha +2 \partial_\mu \alpha \partial^\mu \beta) + \dfrac{\alpha^2 - \Mpd}{\Mp}\partial_\mu \alpha \partial^\mu \gamma - (\alpha^2+\Mpd) \left( \bar{\Box}\alpha + \bar{\Box}\beta + \dfrac{\alpha}{\Mp} \bar\Box\gamma\right) + \dfrac{2(\alpha^2+\Mpd)^2}{3\Mpd}V_{,\beta}=0\label{5.5b}\\
&\gamma: \hspace{0.3cm} -2\alpha \partial_\mu \alpha \partial^\mu \gamma + \dfrac{\alpha^2 - \Mpd}{\Mp}(\partial_\mu \alpha \partial^\mu \alpha + \partial_\mu \alpha \partial^\mu \beta) - (\alpha^2 + \Mpd)  \left( \dfrac{\alpha}{\Mp}(\bar\Box \alpha + \bar\Box \beta) - \bar\Box \gamma\right) + \dfrac{2(\alpha^2+\Mpd)^2}{3\Mpd} V_{,\gamma}=0\label{5.5c} \\
& g_{\mu \nu}: \hspace{0.3cm} R_{\mu \nu} - \dfrac{1}{2}g_{\mu \nu} R = \dfrac{1}{\Mpd} T_{\mu \nu}\label{5.5d}
\end{align}
\end{subequations}
where the stress-energy tensor is
    $$T_{\mu \nu} = \dfrac{3\Mpd}{2(\alpha^2+\Mpd)} \left[ (\partial_\mu \alpha +\partial_\mu \beta)(\partial_\nu \alpha + \partial_\nu \beta) + \dfrac{2\alpha}{\Mp}(\partial_\mu \alpha + \partial_\mu \beta) \partial_\nu \gamma - \partial_\mu \gamma \partial_\nu \gamma \right] $$ 
    \begin{equation}\label{5.60}
    - g_{\mu \nu} \left[ \dfrac{3\Mpd}{4(\alpha^2+\Mpd)} \left( (\partial_\lambda \alpha +\partial_\lambda \beta)(\partial^\lambda \alpha + \partial^\lambda \beta) + \dfrac{2\alpha}{\Mp}(\partial_\lambda \alpha + \partial_\lambda \beta) \partial^\lambda \gamma  - \partial_\lambda \gamma \partial^\lambda\gamma\right) + V \right].
\end{equation}
The simultaneous presence of $\alpha$ and $\beta$ has an ambiguous meaning because both of them in absence of torsion or simply when they take a constant value, e.g. settling down at a minimum of the potential, play the same role as the Immirzi parameter. In other words, it is not clear which of the two or their sum is the correct dynamical field that eventually falls into the Immirzi parameter of LQG. An important remark on the EOM \eqref{5.5} is the following. Studying inflation, a typical assumption in a single scalar field theory is that the inflaton moves very slowly down to a minimum of the potential, thus neglecting the acceleration with respect to the friction term and solve the system for different choices of the potential. This can also be imposed in our case. For instead take the case $\gamma=0$ in flat FLRW as we will treat below. Due to the similar nature $\alpha$ and $\beta$ and their interaction, it is possible to sum \eqref{5.5a} and \eqref{5.5b} to get rid of the acceleration and friction terms, simplifying even further. In some cases it is possible to obtain an analytic solution for the scale factor and the fields. Subsequently, in addition to $\gamma=0$, we shall investigate the case $\alpha=0$ and see that it corresponds to a Quintom theory \citep{Quin1} \citep{Quin2}.

\section{Cosmological models} \label{Sec.4}
In the following we will work with the spatially flat FLRW metric:
$$ ds^2 = -dt^2 + a(t) \left( dr^2 + r^2 d\theta^2 + r^2 \sin^2(\theta) d\varphi^2 \right)$$
and assume that all the fields are non stationary, homogeneous and isotropic. With these properties, \crefrange{5.5a}{5.5d} become
\begin{subequations} \label{5.6}\small
    \begin{align}
 &\alpha: \hspace{0.2cm} \alpha\left((\alp)^2 -(\bp)^2 + (\gp)^2 \right) - \dfrac{\alpha^2 - \Mpd}{\Mp} \bp \gp - (\alpha^2+\Mpd) \left[ \als + \bs + 3H (\alp +\bp) + \dfrac{\alpha}{\Mp}(\gs +3 H \gp) \right] - \dfrac{2(\alpha^2+\Mpd)^2}{3\Mpd} V_{,\alpha}=0 \label{5.6a} \\
 &\beta: \hspace{0.2cm}  -2\alpha \alp ( \alp + \bp) - \dfrac{\alpha^2 - \Mpd}{\Mp} \alp \gp + (\alpha^2+ \Mpd) \left[ \als + \bs + 3H (\alp +\bp) + \dfrac{\alpha}{\Mp}(\gs + 3H \gp) \right] + \dfrac{2(\alpha^2+\Mpd)^2}{3\Mpd}V_{,\beta}=0\label{5.6b}\\
&\gamma: \hspace{0.2cm} 2\alpha \alp \gp - \dfrac{\alpha^2 - \Mpd}{\Mp} \alp (\alp + \bp) + (\alpha^2+ \Mpd) \left[ \dfrac{\alpha}{\Mp} ( \als + \bs + 3H (\alp +\bp))- (\gs +3H\gp) \right] + \dfrac{2(\alpha^2+\Mpd)^2}{3\Mpd} V_{,\gamma}=0\label{5.6c} \\
& \textit{tt}: \hspace{0.2cm} 3H^2 - \dfrac{3}{4(\alpha^2+\Mpd)} \left[ (\alp + \bp)^2 + \dfrac{2\alpha}{\Mp}(\alp + \bp)\gp -(\gp)^2 \right] -\dfrac{1}{\Mpd}  V =0\label{5.6d} \\
& \textit{rr}:\hspace{0.2cm}  -2 \dot{H} -3 H^2 - \dfrac{3}{4(\alpha^2+\Mpd)} \left[ (\alp + \bp)^2 + \dfrac{2\alpha}{\Mp}(\alp + \bp)\gp -(\gp)^2 \right] + \dfrac{1}{\Mpd}  V =0\label{5.6e}
\end{align}
\end{subequations}
\noindent
where $H= \apr /a$ is the Hubble function, $tt$ corresponds to the first Friedmann equation and \eqref{5.6e} is the $rr$ component of the EFE. From \eqref{5.6d} and \eqref{5.6e} we see that the energy density and pressure are respectively given by
$$ \rho = \dfrac{3\Mpd}{4(\alpha^2+\Mpd)} \left[ (\alp + \bp)^2 + \dfrac{2\alpha}{\Mp}(\alp + \bp)\gp -(\gp)^2 \right] + V$$
$$ p = \dfrac{3\Mpd}{4(\alpha^2+\Mpd)} \left[ (\alp + \bp)^2 + \dfrac{2\alpha}{\Mp}(\alp + \bp)\gp -(\gp)^2 \right] - V.$$ 

\begin{center}
\subsection*{Case \texorpdfstring{$\gamma=0$}{Lg}}
\end{center}
This choice corresponds to the action used in LQG with the suitable difference that instead of the Immirzi parameter attached to both the Holst and the Nieh-Yan terms we have two interacting scalar fields subjected to the influence of the potential. Our main strategy is to obtain the dynamics of the universe in the very early time when one or more scalar fields are supposed to be responsible for the inflation era and the corresponding evolution of the fields. In this picture $\alpha$ will be thought as the inflaton while $\beta$ as a potential free scalar. According to the standard inflation theory, at the end of inflation $\alpha$ gets closer to the minimum, oscillates around it giving rise to reheating from which the Hot Big Bang follows. With this, $\alpha$ becomes very nearly constant (Immirzi parameter) and the Holst extra piece can be regarded at least classically as a topological invariant. However, we will also study some specific potentials leading to diverse albeit noteworthy consequences. For the matter in question, the system \eqref{5.6} reduces to 
\begin{subequations}\label{5.7}
    \begin{align}
 &\alpha + \beta: \hspace{0.2cm} \dfrac{3\Mpd}{2(\alpha^2+\Mpd)}(\alp + \bp)^2 + \dfrac{\alpha^2+\Mpd}{\alpha} (V_{,\alpha} - V_{,\beta} )=0 \label{5.7a} \\
 & \textit{tt} + \textit{rr}:\hspace{0.2cm}  \dot{H} + \dfrac{3}{4(\alpha^2+\Mpd)} (\alp + \bp)^2=0.\label{5.7b}\\
& \textit{tt}: \hspace{0.2cm} 3H^2 - \dfrac{3}{4(\alpha^2+\Mpd)} (\alp + \bp)^2 -\dfrac{1}{\Mpd}  V =0\label{5.7c} 
\end{align}
\end{subequations}
with energy density and pressure
\begin{equation}\label{5.8}
     \rho = \dfrac{3\Mpd}{4(\alpha^2+\Mpd)} ( \alp + \bp)^2 + V
     \end{equation}
\begin{equation}\label{5.9}
     p = \dfrac{3\Mpd}{4(\alpha^2+\Mpd)}(\alp + \bp )^2 - V
\end{equation}
In the framework of inflation the potential dominates over the kinetic energy, namely
$$\dfrac{3\Mpd}{2(\alpha^2+\Mpd)}(\alp + \bp )^2 \ll V$$
from which the first Friedmann equation \eqref{5.7c} can be approximated as:
\begin{equation}\label{6.0}
  H^2 \approx \dfrac{1}{3\Mpd}  V.
  \end{equation}
  The acceleration of the scale factor translates as
  \begin{equation}\label{6.1}
      \dfrac{\as}{a}=(1-\epsilon)H^2>0
  \end{equation}
  giving the slow-roll parameter
  $$\epsilon = -\dfrac{\dot{H}}{H^2} \ll 1.$$
  In our case, without using the approximation \eqref{6.0}, we can exactly find it from \eqref{5.7}:
  \begin{equation}\label{6.2}
      \epsilon = \dfrac{3}{1- \dfrac{2\alpha}{\alpha^2+\Mpd}\dfrac{V}{V_{,\beta} - V_{,\alpha}}} \ll 1 .
  \end{equation}
  At first sight this parameter could be complex if $\alpha$ is imaginary. However, since the Hubble parameter must be real, so is its time derivative. By \eqref{5.7b} we see that the second term on the left-hand side of \eqref{5.7a} has to be real and this ensures that \eqref{6.2} is real.
\subsubsection*{Example 1}
\begin{center}
\fbox{$V(\alpha)=-V_0 \; ln\left(\dfrac{\alpha^2}{\Mpd} +1\right), \hspace{1cm} V_0\in \mathbb{R}$}
\end{center}
Such a potential is real and the constant $V_0$ has dimension $[M]^4$. The field $\beta$ is not subjected to its own potential but only to the evolution of $\alpha$. From \eqref{5.7a} we get
\begin{equation}\label{6.3}
    \dfrac{3\Mpd}{4(\alpha^2+\Mpd)}(\alp + \bp)^2= V_0,
\end{equation}
that plugged into \eqref{5.7b}:
\begin{equation}\label{6.4}
    \dot{H} + \dfrac{V_0}{\Mpd}=0
\end{equation}
leads to the solution for the scale factor
\begin{equation}\label{6.5}
        a(t) = a_0 e^{-\frac{V_0}{2\Mpd}t^2+ct}
\end{equation}
where $a_0,c \in \mathbb{R}$ and $H(t) = -V_0t/\Mpd+c$. The exponential expansion is the same as in chaotic inflation  \citep{CI2} \citep{CI1} \citep{Riotto} provided $V_0>0$. However we expect that the time evolution of $\alpha$ will not be linear. The comoving Hubble sphere $(aH)^{-1}$ decreases in the inflation era, which translates into a strictly positive acceleration of the scale factor:
$$\ddot{a}(t) = \left( H^2(t) - \dfrac{V_0}{\Mpd} \right)a(t)>0 \hspace{0.5cm} \iff \hspace{0.5cm} -\infty<t< \dfrac{\Mpd c}{V_0} - \dfrac{\Mp}{\sqrt{V_0}}.$$
According to this scenario, inflation starts back in time at infinity and stops at $t_{fin}=\dfrac{\Mpd c}{V_0} - \dfrac{\Mp}{\sqrt{V_0}}$. We now solve \eqref{5.7c} for the field $\alpha$:
\begin{equation}\label{6.6}
    \alpha(t) = \pm \Mp \sqrt{e^{1-\frac{3\Mpd}{V_0}H^2(t)}-1}
\end{equation}
Going back to infinity the field is purely imaginary 
\begin{equation}\label{6.7}
    \displaystyle{\lim_{t\rightarrow -\infty}} \alpha(t) = \pm i \Mp
\end{equation} 
or alternatively the unrescaled dimensionless scalar field is just $\pm i$. Since the field is imaginary the potential is convex and $\alpha$, starting from a value close to the Planck mass, rolls down to the global minimum located at $\alpha=0$. For large-field models it is believed that the universe emerged from a state where the energy density is comparable to that of the Planck energy density: $\rho \approx \Mp^4$, therefore we can find the initial value of the field $\alpha_{in}$ when $\rho(\alpha_{in}) \approx \Mp^4$. It gives
\begin{equation}\label{6.12}
    \alpha_{in} \approx \pm \Mp \sqrt{e^{1-\frac{\Mp^4}{V_0}}-1}.
    \end{equation}
    To obtain the order of magnitude of $V_0$ we first find the initial value of the Hubble parameter $H_{in} :=  H(t_{in})$ by comparing \eqref{6.6} and \eqref{6.12}:
    \begin{equation}\label{6.13}
        H_{in}^2 = \dfrac{\Mpd}{3}
    \end{equation}
    and impose that the number of e-folds generated during inflation is about 60:
    \begin{equation} \label{6.12+}
    N = \int_{t_{in}}^{t_{fin}} dt' \, H(t') = \int_{H_{in}}^{H_{fin}} dH'  \, \dfrac{H'}{\dot{H'}} = \dfrac{\Mpd}{2V_0} \left( H_{in}^2 - H_{fin}^2\right)\approx 60
    \end{equation}
    that yields
    \begin{equation}\label{6.14}
        V_0 \approx \dfrac{\Mp^4}{360}.
    \end{equation}
    Thus, inflation occurs in the region $\Mp^4 \lesssim \rho \lesssim 10^{-2} \Mp^4$. At the beginning and the end of inflation, the inflaton takes the values
    \begin{equation}\label{6.15}
    \alpha(t_{in}) \approx \pm \Mp \sqrt{e^{-360}-1}
\end{equation}
\begin{equation}\label{6.8}
    \alpha(t_{fin}) = \pm i \Mp \dfrac{\sqrt{e^2-1}}{e}
\end{equation}
It is also worth knowing the evolution of the sum of the fields. Taking \eqref{6.3} and using \eqref{6.6} it gives
\begin{equation}\label{6.9}
    (\alpha + \beta) (t) = \pm \Mp \dfrac{\sqrt{2\pi e}}{3} erf \left( -\sqrt{\dfrac{3}{2}}\dfrac{\Mp}{\sqrt{V_0}} H(t) \right) + const.
\end{equation}
where $erf$ is the error function. First, as expected, their sum is real. Secondly, since the Hubble function is slowly varying during inflation, so is \eqref{6.9}. Indeed
\begin{equation}\label{6.10}
    \left( \alpha + \beta \right) (t_{in}) = \pm \Mp \dfrac{\sqrt{2\pi e}}{3} + const
\end{equation}
while at the end of inflation
\begin{equation}\label{6.11}
   (\alpha + \beta) (t_{fin}) = \pm \Mp \dfrac{\sqrt{2\pi e}}{3}erf\left(\sqrt{\dfrac{3}{2}}\right) + const.
\end{equation}
The value $w = p/\rho$ , using \eqref{5.8}, \eqref{5.9} and \eqref{6.3}, \eqref{6.6} reads
\begin{equation}\label{6.16}
    w = -1 + \dfrac{2V_0}{3\Mpd H^2}
\end{equation}
therefore $-1<w<-1/3$ and there is no phantom phase. After a certain time $\alpha$ rolls down, reheates giving birth to particles which shape the features of the perfect fluid. Additional energy density $\rho_{ext}$ and $p_{ext}$, where \textit{ext} stands for external contributions, appear in the Friedmann equations. If $\alpha$ and $\beta$ do not couple to any external fields kinetically and potentially, then \crefrange{5.7a}{5.7c} become
\begin{subequations}\label{6.17}
\begin{align}
&\dfrac{3\Mpd}{2(\alpha^2+\Mpd)}(\alp + \bp)^2 + \dfrac{\alpha^2+\Mpd}{\alpha} V_{,\alpha}=0 \label{6.17a} \\
&\dot{H} + \dfrac{3}{4(\alpha^2+\Mpd)} (\alp + \bp)^2 + \dfrac{\rho_{ext}(1+w)}{2\Mpd}=0 \label{6.17b} \\
&3H^2 = \dfrac{3}{4(\alpha^2+\Mpd)} (\alp + \bp)^2 + \dfrac{1}{\Mpd}  V(\alpha) + \dfrac{\rho_{ext}}{\Mpd}\label{6.17c}
\end{align}
\end{subequations}
where in \eqref{6.17b} we used $p_{ext}= w \rho_{ext}$. Being of non-phantom nature, the energy density of the interacting pair 
$$\rho = \dfrac{3\Mpd}{4(\alpha^2+\Mpd)} (\alp + \bp)^2 +  V(\alpha)= V_0\left(1 - ln\left(\dfrac{\alpha^2}{\Mpd}+1\right)\right)$$
dilutes as the universe enlarges which implies that $\alpha$ has to roll down the potential because the kinetic energy is constant. It goes down up to a point when $\left|\alpha^2\right| /\Mpd \ll 1$, so that we can expand the potential around the minimum:
$$V(\alpha) = - \dfrac{V_0}{\Mpd} \alpha^2 + O(\alpha^4)$$
hence \eqref{6.17c} becomes
\begin{equation}\label{6.18}
    3H^2 = \dfrac{V_0}{\Mpd}(1-\dfrac{\alpha^2}{\Mpd})+ \dfrac{\rho_{ext}}{\Mpd}
\end{equation}
namely we ended up with the first Friedmann equation with a varying "cosmological constant" in natural units which asymptotically tends to a constant value. Unlike a common idea that the cosmological constant might be the value of the potential of a scalar field at which it sits, in our case for sufficient long time its limit is instead the value of the kinetic energy. As $\alpha$ slows down, $\beta$ speeds up to compensate, breaking free and behaving like a free scalar field. It is worth mentioning that this result holds independently from the type of matter content of the universe. However, there is a problem with this result. The cosmological constant we obtained tends to a constant value that is far away from the one we measure by observations. This issue stems from the assumption that the initial energy density has to be of order of the Planck density, which implies that $V_0$ has to be enormous and the Hubble constant is about the Planck mass. If instead we drop that condition and allow $V_0$ to be very small, then $H$ has to be extremely small too in order to have $N \approx 60$ hinting at an universe born from vacuum fluctuations. That is, if we set $V_0/\Mpd = \Lambda$, the condition \eqref{6.12+} yields $H_{in}^2 = 121 \Lambda$ and $H_{fin}^2 = \Lambda$. Working in SI, $V_0 =\Lambda/(\kappa c^2) \approx 5.5 \, 10^{-27} kg/m^3$ and following the same procedure as above $\rho_{in}c^2 \approx 38 \, GeV/m^3$ is the initial energy density.  It should be carried out a perturbation analysis to compute the parameters needed to compare the model to observations but we stop here leaving this task to future works.
\subsubsection*{Example 2}
\begin{figure}[h]
    \includegraphics[height=5.5cm]{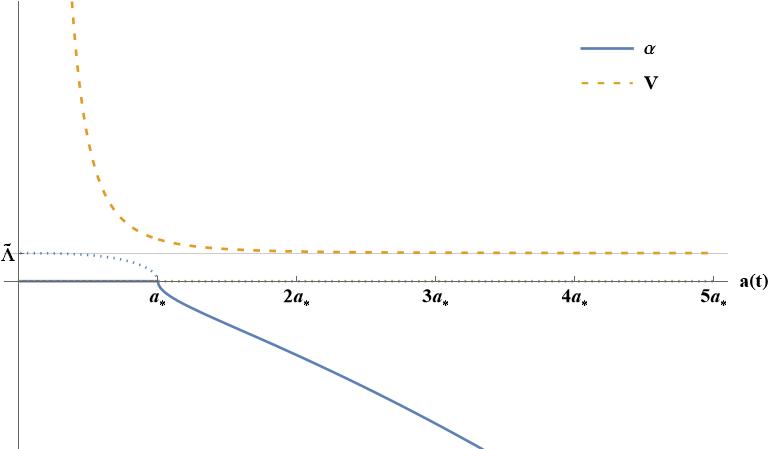}
    \caption{Display of the evolution of $\alpha$ and $V$ with respect to the scale factor. The dotted line corresponds to the imaginary part of $\alpha$, the thick solid line its real part and the dashed line the potential (real). The value at which the field switches is $a_\star = \left(\Lambda \Mpd \Omega_m/ 2V_{-1} \Omega_\Lambda\right)^{1/3}$ after which it runs forward at a rate $\propto a(t)^{3/2}$. The potential, on the other hand decreases fast $\propto a(t)^{-3}$, eventually approaching $\tilde{\Lambda}$ (the thin solid line represents only the asymptotic value of the potential).}
    \label{fig.pot}
\end{figure}

\begin{center}
\fbox{ $V=V_n\left(\dfrac{\alpha^2}{\Mpd} + 1\right)^n+ \tilde{\Lambda} \;, \hspace{1cm} (V_n,\tilde{\Lambda}) \in \mathbb{R}^2$}
\end{center}
where $n \in \mathbb{Z}$.
Eq. \eqref{5.7a} yields
\begin{equation}\label{8.0}
    \dfrac{3}{4(\alpha^2 + \Mpd)}(\alp+\bp)^2= -\dfrac{nV_n}{\Mp^{2n+2}}(\alpha^2 + \Mpd)^n
\end{equation}
that put into \eqref{5.7b} gives
\begin{equation}\label{8.1}
    \dot{H} = \dfrac{nV_n}{\Mp^{2n+2}}(\alpha^2+\Mpd)^n
\end{equation}
while \eqref{5.7c}, using \eqref{8.1}, becomes
\begin{equation}\label{8.2}
     (n-1) \dfrac{\as}{a} + (2n+1) \left(\dfrac{\apr}{a}\right)^2 - \dfrac{n\tilde{\Lambda}}{\Mpd} =0. 
\end{equation}
that produces simple solutions for the scale factor:
\begin{subequations}\label{8.3}
\begin{align}
     &a_{+}(t) = k_0 \sinh^{\frac{n-1}{3n}}{\left(At\right)} & \tilde{\Lambda}>0 \label{8.3a} \\ 
     &a_0(t) = k_0 t^{\frac{n-1}{3n}} & \tilde{\Lambda}=0 \label{8.3b} \\
     &a_{-}(t) = k_0 \sin^{\frac{n-1}{3n}}{\left(At\right)} &\tilde{\Lambda}<0 \label{8.3c}
     \end{align}
\end{subequations}
with $A=n\sqrt{3\vert \tilde{\Lambda}\vert}/(n-1)\Mp$. The eq.\eqref{8.3b} can also be recast in the form of $a_0(t)=k_0 t^{2/3(w+1)}$ with the equation state parameter $w=(n+1)/(n-1)$. Thus, $n=-1$ mimics a matter-dominated universe, whereas $n=-2$ a radiation-dominated one. The presence of $\tilde{\Lambda}$  can be seen as the cosmological constant which exponentially blows \eqref{8.3a} in time and oscillates \eqref{8.3c} depending on its sign. In the $\Lambda CDM$ model a matter+cosmological constant dominated universe provides \citep{LCDM} $a(t) = (\Omega_m/\Omega_\Lambda)^{1/3}\sinh^{2/3}{(3H_0\sqrt{\Omega_\Lambda}t/2)}$, where $\Omega_m$ and $\Omega_\Lambda$ are the density parameters of matter and dark energy respectively. For $ \tilde{\Lambda}>0$,  $n=-1$, $k_0=(\Omega_m/\Omega_\Lambda)^{1/3}$ and $A=3H_0\sqrt{\Omega_\Lambda}/2$ in \eqref{8.3a}, we obtain $\tilde{\Lambda}=3 \Mpd\Omega_\Lambda H_0^2 = \Mpd \Lambda$, which is the cosmological constant (up to $\Mpd$) expressed in natural units\footnote{Remind that the cosmological constant has dimension $[L]^{-2}$ and in the EH action is multiplied by $\Mpd$, whereas $\tilde{\Lambda}$ has dimension $[L]^{-4}$, hence the extra multiplicative factor $\Mpd$.}. Solving \eqref{8.1} for $\alpha$ it gives
\begin{equation}\label{8.4}
    \alpha(t)= \pm i \Mp \sqrt{\left( \dfrac{V_n(n-1)\sinh^2\left(At\right)}{\tilde{\Lambda}}\right)^{-1/n}+1}
\end{equation}
Let us study the most interesting case $n=-1$. Then
\begin{equation}\label{8.5}
    \alpha(t)= \pm i \Mp \sqrt{1- \dfrac{2 
 V_{-1}}{\Mpd \Lambda}\sinh^2\left(\dfrac{\sqrt{3\Lambda}}{2}t\right)}
\end{equation}
and using \eqref{8.3a}
\begin{equation}\label{8.6}
  \alpha(a(t))= \pm i \Mp \sqrt{1- \dfrac{2 
 V_{-1}\Omega_\Lambda}{\Mpd \Lambda\Omega_m}  a^3(t)}
 \end{equation}
Such a field can be either real or imaginary. For $V_{-1}<0$, $\alpha$ is imaginary and cannot cross the barrier $\pm i \Mp$ toward zero and must be rejected. More compelling is the case $V_ {-1}>0$ for which the field starts being imaginary for $0<a(t)<\left(\Lambda \Mpd \Omega_m/ 2V_{-1} \Omega_\Lambda\right)^{1/3}$ and turns real afterward. It is surprising that it depends on the constant $V_{-1}$ while the potential does not change with respect to its coefficient:
\begin{equation}\label{8.7}
    V(a(t)) = \Mpd \Lambda \left( 1 + \dfrac{\Omega_m}{2\Omega_\Lambda a^3(t)}\right).
\end{equation}
From \eqref{8.6} and \eqref{8.7} the scalar field increases over time falling along the potential that flattens down to the value of $\tilde{\Lambda}$, tending to a De-Sitter universe (see \figurename{\ref{fig.pot}}).
 \\ \noindent
The evolution of the pair can be easily extracted from \eqref{8.0}:
\begin{equation}\label{8.8}
    (\alpha+\beta)(t) = 2\sqrt{\dfrac{V_{-1}}{3}}\:t+const.
\end{equation}
If the Barbero-Immirzi field is set to be\footnote{Here $\gamma$ stands for the Immirzi character used in LQG} $\gamma(t) := 1/((\alpha + \beta)(t))$, it asymptotically converges to zero. Both the first and second example are the only nontrivial analytical solutions and coincidentally exhibit one of the most studied theory of inflation and the most likely course of the Cosmos after inflation, respectively. Although it could be possible to sum up the potentials of the previous example and the ones with $n=-2$ and $n=-1$ of this case and arrange the parameters in such a way to have a fluent evolution in which each potential dominates for a span of time, it is na\"ive to think that scalar fields can be enough to describe the stages of the universe without taking into account any other content filling it. Secondly, due to the very structure of the equations, $\alpha$ is not found from a differential equation, therefore no constant of integration appears to freely choose its initial value. Conversely, the solution of $\beta$ involves a first order differential equation and we always have a constant emerging from it.

\subsubsection*{Example 3}
\begin{figure}[h]
    \subfloat{\includegraphics[height=4.7cm]{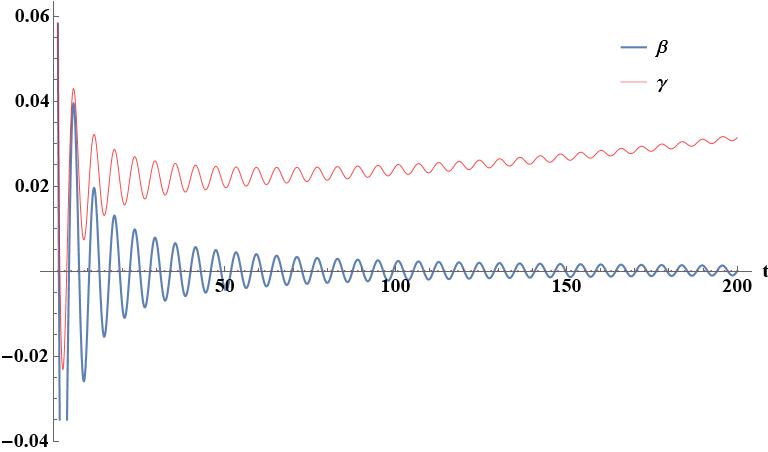}}
    \qquad
    \subfloat{\includegraphics[height=4.7cm]{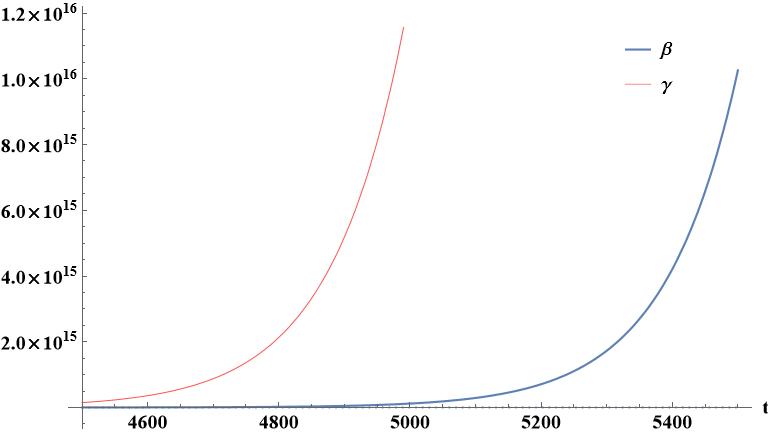}}
     \caption{Evolution of $(\beta,\gamma)$ during a matter dominated universe. The thick blue line draws $\beta$ and the thinner red line $\gamma$. We set the values $M=1.30$, $m=0.013$, $g=0.094$, $c_1=0.23$, $c_2=2.44$. The inflaton keeps swinging around zero because of its heavy mass, whereas the phantom field keeps oscillating around an up-advancing point as the friction term decreases over time (left-hand side graph). Eventually the phantom field blows first, carrying the other with it (right-hand side graph), breaking the matter dominance.} \label{fig.beta-gamma}
\end{figure}

Starting from \eqref{5.7a}, if we consider the scalar fields as a single entity in the potential, namely if $V(\alpha,\beta)=V(\alpha+\beta)$, whatever the shape of the potential is, or just a constant, then $V_{,\alpha} - V_{,\beta} =0$ and for $\alpha \neq 0$ the only solution is
\begin{equation}\label{9.0}
    \alpha(t) + \beta(t) = const.
\end{equation}
Thus, if the Barbero-Immirzi field is taken to be the sum of them, it is just an arbitrary parameter of the theory in accordance with LQG. In absence of other sources, the potential plays the role of the cosmological constant. In other words, in this simple model, $(\alpha,\beta)$ can be thought as a bound pair in which the fields counter each other making them ineffectual in influencing gravity. In the case of \eqref{6.17}, being \eqref{6.17a} zero, the kinetic terms in \eqref{6.17b} and \eqref{6.17c} vanish and we are left with matter fields shaping gravity in presence of a cosmological constant. Adding by hand external energy and pressure of a perfect fluid engender different outcomes than considering them from the beginning in first order formalism. For instead, if in the action \eqref{5.0} we append the Lagrangian of massive fermions, as known, the fermion current $J^\mu$ is attached to the pseudotrace axial vector \citep{Nasci} likewise $\alpha$ and $\beta$. Thus, solving the EOM, all these fields are kinematically connected, eq.\eqref{9.0} no longer holds and if the potential does not contain interaction between scalars and fermions, $\alpha,\beta$ would not be isotropic  and would depend on the fermion current. In the hypothesis that only scalar fields fill the very early universe and matter is created afterward, it is questionable whether the first order formalism has to be solved with both of them. \noindent
\begin{center}
\subsection*{Case \texorpdfstring{$\alpha=0$}{Lg}}
\end{center}
Setting $\alpha=0$, the initial action we start from is
$$ S= \dfrac{1}{2}\int \Bigl( \Mpd e^I \wedge e^J \wedge \star R_{JI}+  \Mp \beta(x) d(e^I \wedge T_I) + \Mp \gamma(x) d(e^I \wedge \star T_I) \Bigr)$$
namely including the only two closed four-forms that can be constructed from the torsion. The effective action is 
\begin{equation}\label{10.0}
\small
    S= \dfrac{\Mpd}{2}\int_{\mathcal{M}} d^4x \; e\;  e^\mu_I e^\nu_J \bar{R} \I{_\mu_\nu^I^J} - \dfrac{3}{4} \int_{\mathcal{M}} d^4x \; e\; \left( \partial_\mu\beta \partial^\mu \beta - \partial_\mu \gamma \partial^\mu \gamma\right).
\end{equation} 
where we see that the absorption of $\Mp$ by the fields makes the action canonical. In this framework the Immirzi field is played by $\beta$ while $\gamma$ is a phantom field. Following \citep{Quin2}, if $\beta,\gamma$ are real scalar fields, they can be thought as the real and imaginary part of a complex field $\phi = \beta + i \gamma$ whose non-orthodox action is
$$ S = -\dfrac{3}{8}\int_{\mathcal{M}} d^4x \; \sqrt{-g}\; \left( \partial_\mu \phi \partial^\mu \phi + \partial_\mu \phi^\star \partial^\mu \phi^\star\right). $$
Defining $\Phi^T = (\phi,\phi^\star)$, the bracket of such a Lagrangian can be written as $\partial_\mu \Phi^T \partial^\mu \Phi$, therefore it is invariant under the global group $O(2,\mathbb{C})$ which is abelian, isomorphic to $\mathbb{C}^\star$ and its Lie Algebra is simple, semisimple and non-compact. For further reading see also \citep{Quin3}. Going back to our task the EOMs \eqref{5.6} reduce to 
\begin{subequations} \label{10.1}
    \begin{align}
 &\bs + 3H\bp + \dfrac{2}{3} V_{,\beta}=0\label{10.1a} \\
&\gs + 3H\gp - \dfrac{2}{3} V_{,\gamma}=0\label{10.1b} \\
&3H^2 - \dfrac{3}{4 \Mpd}  (\bp^2 -\gp^2) -\dfrac{1}{\Mpd} V =0\label{10.1c} \\
& \dot{H} + \dfrac{3}{4\Mpd} (\bp^2 -\gp^2)=0 .\label{10.1d}
\end{align}
\end{subequations}
Usually a way to treat such a theory is to view the field $\beta$ as the driving entity responsible for inflation and $\gamma$ as the phantom that at late times guides the universe into a fast expansion with $w(t) \leq -1$. If $w$ is constant at a finite time the scale factor blows up to infinity \citep{Doom} \citep{PhantomCosmologies}. If it varies there are two scenarios: for bounded potentials the resulting attractor is De-Sitter, for unbounded potential it could lead to a De-Sitter universe, a Big Rip or take a different shape, depending on the steepness and the form of the potential \citep{Faraoni} \citep{Sami} \citep{Odintsov}. Concisely, once the inflaton reheates the cosmos passes through different eras with equation of state depending on its content, while the phantom field remains almost still due to its nearly thoroughly flat potential, eventually overwhelming all other contributions that fill a dilating universe. In order for inflation to be consistent we require that the friction terms in \eqref{10.1a} \eqref{10.1b} outweigh the acceleration of both fields and the slow-roll approximation ensures that the potential dominates over the kinetics. Thus, 
\begin{subequations} \label{10.2}
    \begin{align}
 &3H\bp + \dfrac{2}{3} V_{,\beta} \approx 0\label{10.2a} \\
&3H\gp - \dfrac{2}{3} V_{,\gamma} \approx 0\label{10.2b} \\
&3H^2 \approx \dfrac{1}{\Mpd} V\label{10.2c} \\
& \dot{H} + \dfrac{3}{4\Mpd} (\bp^2 -\gp^2)=0 .\label{10.2d}
\end{align}
\end{subequations}
We choose a potential which includes a quadratic interaction to be 
\begin{equation}\label{10.3}
    V = \dfrac{1}{2}M^2 \beta^2 + \dfrac{1}{2}m^2 \gamma^2 - g^2 \beta \gamma
\end{equation}
where $M$ is the inflaton mass, $m$ is the phantom mass and $g$ is the coupling constant of mass dimensions. Such a potential is unbounded from below, however, for $g^2<M m$ it is strictly positive. We will work in this regime. Normally the phantom mass is considered extremely small ($10^{-30} \,-\, 10^{-60}$) because without interactions its effect in the early universe is negligible and activates only at late times when its energy density exceeds any other one. We work in large-field model for $\beta$ and small values for the phantom, possibly very close to zero, so that the only contribution to inflation is the first term in \eqref{10.3}. The system \eqref{10.2} with the choice of \eqref{10.3} and the approximations claimed above becomes
\begin{subequations} \label{10.4}
    \begin{align}
 &3H\bp + \dfrac{2}{3} M^2 \beta \approx 0\label{10.4a} \\
&3H\gp + \dfrac{2}{3} g^2 \beta \approx 0\label{10.4b} \\
&3H^2 \approx \dfrac{1}{\Mpd} M^2 \beta\label{10.4c} 
\end{align}
\end{subequations}
 yielding the solution 
\begin{subequations}\label{10.5}
\begin{align} 
    &\beta(t) = \beta_0 - \left(\dfrac{2}{3}\right)^{3/2} \Mp M t \label{10.5a} \\
    & a(t) = a_0 \: e^{ -\frac{M^2}{9} \,t^2 + \frac{M \beta_0}{\sqrt{6}\Mp} \,t}\label{10.5b} \\
    & \gamma(t) = \gamma_0 - \left(\dfrac{2}{3}\right)^{3/2}  \dfrac{g^2 \Mp}{M} t \label{10.5c}
    \end{align}
\end{subequations}
The phantom does not affect inflation, but falls deeper down as $\beta$ diminishes. The reason is that although the phantom tends to run up, initially, being located around zero, feels the influence of the falling of the inflaton and the force $F_\gamma = -V_{,\gamma}$ in \eqref{10.4b} pushes it downward because of the positive sign. Moreover, if $m$ is smaller than $M$, it moves toward zero at a lower speed compared to \eqref{10.5a} due to the coupling constant and the ratio $\Mp/M$. Once the universe stops accelerating, with the creation of matter, both the fields oscillate around the minimum of the potential for a lapse of time. When the vast majority of energy density is composed by the matter one $\rho_m$, we would like to study what occurs to the pair $(\beta, \gamma)$. In such a time gap, \crefrange{10.1a}{10.1c} result
\begin{subequations} \label{10.6}
    \begin{align}
 &\bs + 3H\bp + \dfrac{2}{3}(M^2 \beta -g^2 \gamma)=0\label{10.6a} \\
&\gs + 3H\gp - \dfrac{2}{3} (m^2 \gamma -g^2 \beta)=0\label{10.6b} \\
&3H^2 \approx \dfrac{\rho_m}{\Mpd} \label{10.6c}
\end{align}
\end{subequations}
which, with a suitable choice of the integration constants, \eqref{10.6a}, \eqref{10.6b} yield
\begin{subequations} \small\label{10.7}
    \begin{align}
 &\beta(t) = \dfrac{1}{t} \left( c_1 \, cosh \left(\dfrac{\sqrt{m^2 -M^2 - \sqrt{(M^2+ m^2)^2 -4g^4}}}{3} \,t\right) +g^2 \, c_2 \, cosh\left(\dfrac{\sqrt{m^2 -M^2 + \sqrt{(M^2+ m^2)^2 -4g^4}}}{3} \,t\right)\right) \label{10.7a} \\ 
 &\gamma(t) = \dfrac{1}{2 g^2 t} \Biggl[ c_1 ( M^2 + m^2 - \sqrt{(M^2 + m^2)^2 - 4 g^4}) \: cosh \left(\dfrac{\sqrt{m^2 -M^2 - \sqrt{(M^2+ m^2)^2 -4g^4}}}{3} \,t\right) \nonumber \\ 
 &+g^2 \, c_2 \, ( M^2 + m^2 + \sqrt{(M^2 + m^2)^2 - 4 g^4}) \: cosh \left(\dfrac{\sqrt{m^2 -M^2 + \sqrt{(M^2+ m^2)^2 -4g^4}}}{3} \,t\right)\Biggr] \label{10.7b}
\end{align}
\end{subequations}
Note that for $g \rightarrow 0$, we recover the dumped oscillation of $\beta$ around the minimum and the exponential growth of $\gamma$. Both of the fields are synchronized, namely they oscillate with the same frequency while slowly growing as depicted in \figurename{\ref{fig.beta-gamma}}).
 \\ \noindent
First, being $g^2<Mm$, then $(M^2+ m^2)^2 -4g^4 \, > \, (M^2+ m^2)^2 -4M^2m^2 =  (M^2- m^2)^2>0$, i.e. both the fields are real. Secondly, the functions of \eqref{10.7a}, \eqref{10.7b} attached to $c_1$ are responsible for the dumped oscillations, while the ones attached to $c_2$ are the cause of the exponential growth. Hence, since $M^2 + m^2 + \sqrt{(M^2 + m^2)^2 - 4 g^4}$ is much bigger than $g^2$, the phantom climbs up first and faster. For example if $M=10^{-6} \Mp$, $m= 10^{-36} \Mp$, $g= 10^{-30} \Mp$, we have
$$\dfrac{M^2 + m^2 + \sqrt{(M^2 + m^2)^2 - 4 g^4}}{g^2} \approx 10^{48}$$ 
and even in the extreme case $M=m=10^{-6} \Mp$, $g= 10^{-10} \Mp$, the ratio is
$$\dfrac{M^2 + m^2 + \sqrt{(M^2 + m^2)^2 - 4 g^4}}{g^2} \approx 10^{8}.$$ 
The reason is straightforward: in \eqref{10.7a} $\beta$ is subjected to a force $M\beta$ that pulls it down and to a force exerted by the coupling that drags it up. Similarly, in \eqref{10.7b}, the mass force pushes $\gamma$ up while the field $\beta$ brings it down. This establishes a frequency between them and make them dance at the same rhythm in an expanding universe. The lighter is the phantom mass the longer it takes for it to climb up and for the pair energy density to take the lead. Once the inflaton-phantom energy density (of which the latter is the biggest responsible) overcomes any other filling, the Cosmos is nearly solely subjected to the dynamics of these two fields. When this occurs, the universe experiences an acceleration state, the Hubble function goes up causing a large friction term and since the two fields are too far away from each other, oscillations are suppressed. Therefore, in the regime of phantom potential preeminence, the system \eqref{10.6} can be approximated as
\begin{subequations} \label{10.8}
    \begin{align}
 & 3H\bp - \dfrac{2}{3} g^2 \gamma \approx 0\label{10.8a} \\
& 3H\gp - \dfrac{2}{3} m^2 \gamma \approx 0\label{10.8b} \\
&3H^2 \approx \dfrac{m^2 \gamma^2}{2\Mpd} \label{10.8c}
\end{align}
\end{subequations} 
that gives
\begin{subequations} \label{10.9}
    \begin{align}
 & \beta(t) = \left( \dfrac{2}{3}\right)^{3/2} \dfrac{g^2 \Mp}{m} \,t +\beta_0^\prime \label{10.9a} \\
& \gamma(t) = \left( \dfrac{2}{3}\right)^{3/2} m \Mp \,t+ \gamma_0^\prime\label{10.9b} \\
& a(t)= a_0^\prime \:e^{ \frac{m^2}{9} \,t^2 + \frac{m \gamma_0^\prime}{\sqrt{6}\Mp} \,t}\label{10.9c}
\end{align}
\end{subequations}
with $\gamma_0^\prime \gg \beta_0^\prime$. Such a behaviour is often called "slow-climb" and the universe undergoes a "reverse inflation". As can be seen from \crefrange{10.9a}{10.9b} if $g<m$, $\beta$ never reaches $\gamma$ and \eqref{10.8} holds forever. The energy density and the pressure
\begin{align}
   & \rho \overset{t \shortrightarrow \infty}{\longrightarrow} +\infty 
   &p \overset{t \shortrightarrow \infty}{\longrightarrow} -\infty
\end{align}
indicate that the Big Rip is inevitable and the spacetime tends asymptotically to De-Sitter, likewise \citep{Sami}. If $g>m$, it is not easy to estimate what could happen. First of all, from \crefrange{10.9a}{10.9b} and \eqref{10.2d} $\dot{H} < 0$, so the friction term decreases over time although it remains remarkable, $\beta$ gets closer to $\gamma$, the contribution of $\beta$ in the potential becomes relevant and oscillations cannot be discarded. It is not clear whether there exists a stable point around which they fluctuate or the fields continue moving up to infinity. \\ \noindent
There is also another aspect to take into account. The energy density could reach the Planck density before \eqref{10.8} break and a quantum theory of gravitation is needed. Another possibility is that if the inflaton achieved the value at which it reheated in the very early universe before the energy density climbs over the Planck density, it would be unclear whether nothing occurs or all the matter that had been created might instantly disappear. A coupling considered here between the inflaton and the phantom is spellbinding because keeps the main features intact: inflation is unchanged, the fields oscillate around the minimum for a long time during a matter dominated universe, the phantom turns on at late times before the inflaton; and mutates the role of the phantom because it drags its counterpart up with it, leaving speculations about what could occur when the inflaton gets back to its reheating point, perhaps an open the door to a cyclic universe.
  \section{Conclusions}
  In this work we considered the picture of using topological invariants as dynamical entities that shape the evolution of the universe in its early and late stages. The idea behind it was to promote, respectively, the coefficients $\alpha, \beta, \gamma$ of the Holst, Nieh-Yan and Nieh-Yan-like invariants to scalar fields in first order formalism. In particular, the first two can be regarded as the Barbero-Immirzi scalar field while the latter was never taken into account in LQG although widely used in Teleparallel Gravity. After detailing some properties of such pieces in Section \ref{Sec.2}, the effective action resulted in a Scalar-Tensor theory of gravity with interacting fields in which a general potential was added (Section \ref{Sec.3}). Its natural implementation in a FLRW universe turned out to be the very complicated system \eqref{5.5} that was simplified by cutting off some d.o.f.: in the first case $\gamma$ was switched off and in the second one $\alpha$. The former corresponds to a kinetically interacting pair  where the primary role is played by $\alpha$ that renders the kinetic term non standard. Despite the very complicated system \eqref{5.6}, it was possible to drastically simplifying it without any assumptions or approximations although the price to be paid was the loss of the canonical form of the EOM for both the fields. Nonetheless, for $\beta$-free potentials analytical solutions were found in three examples. In the first one a logarithmic potential exhibits an inflation scenario that shares some features with chaotic inflation though a remarkable difference lies in the fact that the inflaton $\alpha$ does not roll down linearly but exponentially. After it reheates, matter density enters the game while the inflaton continues its way down to the minimum located at zero. The kinetic energy remains constant while the potential energy diminishes until $\alpha$ becomes so small to be negligible, leaving the kinetic energy enacting the role of the cosmological constant. However, the large-field model was used believing that the universe emerged from a quantum gravity state of Planck regime. In that case the cosmological constant turned out to be too big compared to observations, thus initial low energy density is a more suitable scenario. In the second example a polynomial-like potential was analyzed. Two cases describe a radiation and matter dominated universe with or without cosmological constant. The latter caught our attention. For a strictly positive valued potential the field $\alpha$ starts being imaginary of the order of the Planck mass, going down as the universe expands till it turns real, inverting the trend and begin to increase. As it goes asymptotically to infinity the potential flattens down up to take a value corresponding to the cosmological constant. In the third example a generic type of potential was provided. If it depends on the sum $\alpha + \beta$, their existence does affect gravity as a cosmological constant and more importantly $\alpha + \beta = constant$ that could explain why the Immirzi parameter is indeed constant in LQG. Subsequently, $\alpha$ was switched off in the system \eqref{5.6}, falling into a Quintom theory with $\beta$ a real scalar field and $\gamma$ a phantom. Since in many cases no binding energy is taken into account, namely the potential does not contain interactions between the two fields we studied a specific case with massive fields and a quadratic interaction modulated by a coupling constant. The real field was regarded as the inflaton that drove the opening stage of the universe, swinging around the minimum in damped oscillations while the phantom keeps resting for a long time. The interaction term is such that it can be considered as an "attractive-repelling" force. Indeed, during inflation $\beta$ rolls down, pushing the phantom deeper down. When matter holds the stage the inflaton establishes a relation with the phantom, oscillating at the same pace. If the phantom mass is extremely small, as it thought to be, its preeminence requires a longer time to come up. The energy density of the pair increases rapidly as the phantom is pushed up by its potential, overcoming the previous matter one. The inflaton remains a longer time attached to the minimum with respect to its counterpart until the binding force constrains to chasing the phantom. The Hubble function climbs up making the universe more viscous during which oscillations are suppressed, both the fields move up linearly and the scale factor blows steeply exponentially. If the coupling constant is smaller than the phantom mass, $\gamma$ runs away from $\beta$. On the other hand, if the converse is true, the inflaton could catch the phantom in finite time, complicating the dynamics in a way that no term can be neglected. They could stabilize and fluctuate around a point, so the universe stops enlarging in perpetuity or continue their path to infinity with no way back. We also pointed out that the energy density could exceed the Planck density before a possible equilibrium point is reached and a Quantum Gravity theory is needed. Furthermore, the more interesting point is that the inflaton could surpass the value it took when it reheated, leaving a big question mark about what could happen. Would matter be destroyed in a blink? Would the universe collapse and be born again? Would nothing happen and a Big Rip is inevitable?
  
\section*{Acknowledgments} 
I would like to thank Professor Franz Hinterleitner of Masayk University for precious discussions and consultations. This work was supported by the MUNI/A/1077/2022 Masaryk University grant.
  \medskip
  \bibliographystyle{unsrtnat}
  \bibliography{Immirzi}
  \end{document}